\documentclass[9pt,twocolumn,twoside]{opticajnl}
\journal{opticajournal} 

\setboolean{shortarticle}{true}

\usepackage{siunitx}
\usepackage{braket}
\usepackage{lineno}

\definecolor{blouge}{rgb}{0.5, .1, .6}
\definecolor{bl}{rgb}{0, .1, .6}
\definecolor{turquoise}{rgb}{0.251, 0.878, 0.816}

\usepackage[commentmarkup=uwave]{changes}
\definechangesauthor[color=blue]{dbo}

\title{Probing subradiant dynamics in cold atomic ensembles via population and emitted light measurements}

\author[1]{Antoine Glicenstein}
\author[1,*]{Daniel Benedicto Orenes}
\author[1]{Apoorva Apoorva}
\author[1,**]{Raphaël Saint-Jalm}
\author[1]{Robin Kaiser}

\affil[1]{Université Côte d'Azur, CNRS, Institut de Physique de Nice, 06200 Nice, France}

\affil[*]{daniel.benedicto-orenes@univ-cotedazur.fr}
\affil[**]{raphael.saint-jalm@univ-cotedazur.fr}

\begin{abstract}
In this letter, we report on the time- and space-resolved measurement of subradiant excited state population in an ultra-cold atomic cloud of $^{174}$Yb atoms. We use a \emph{depletion} imaging technique that exploits the \rotatebox[origin=c]{180}{$\Lambda$}-type internal energy structure of alkaline-earth-like atoms to directly observe the time-resolved spatial distribution of excited state population. We characterize the decay dynamics of the subradiant modes using both the excited state population and scattered light intensity, finding good quantitative agreement with numerical predictions from simulations of two-level atomic ensembles.  
\end{abstract}

\setboolean{displaycopyright}{false} 

\begin{document}

\maketitle

Collective scattering of light by atomic ensembles is a well-studied problem in the atomic and quantum optics community. Interactions mediated by real and virtual photons give rise to a wide variety of phenomena, such as Dicke superradiance \cite{Dicke_R._1954,Gross_Haroche_1982}, subradiance \cite{Devoe_Brewer_1996,Pavolini_Crubellier_Pillet_Cabaret_Liberman_1985,Gue15,Ferioli_Glicenstein_Henriet_Ferrier-Barbut_Browaeys_2021}, radiation trapping \cite{Labeyrie_Kaiser_Delande_2005,Weiss_Araújo_Kaiser_Guerin_2018}, modified radiation pressure \cite{Cou10}, or absorption line broadening and shifts \cite{Sut16,Bro16}. These subjects have been studied broadly using ensembles of cold atoms, mainly alkali atoms such as Rb \cite{Ferioli_Glicenstein_Robicheaux_Sutherland_Browaeys_Ferrier-Barbut_2021,Glicenstein_Ferioli_Browaeys_Ferrier-Barbut_2022,Ferioli_Glicenstein_Ferrier-Barbut_Browaeys_2023,Rui_Wei_Rubio-Abadal_Hollerith_Zeiher_Stamper-Kurn_Gross_Bloch_2020,Das2020} or Cs \cite{Pennetta2022,Liedl2024}. In typical experiments, only optical (photon) information is available. As the scattered light is collected outside the atomic ensembles, the internal dynamics must be inferred rather than directly observed. Moreover, a connection between subradiant modes and localized states of light has been predicted \cite{Mor19,Celardo_Angeli_Mattiotti_Kaiser_2024}, which justifies ongoing experimental efforts to observe the transport of light in situ in optically dense atomic ensembles \cite{Gli24}. In this work, we exploit a newly developed imaging technique to measure the excited state population of an ultra-cold atomic ensemble during the subradiant dynamics. The direct detection of excited state population allows access to information about the system which is complementary to the scattered intensity. On the one hand, the scattered intensity is related to the sum of individual atomic coherences \cite{Esp20}:
\begin{equation}
    I(\theta,\phi) = \sum_{i,j} \langle \sigma_i^+ \sigma_j^- \rangle \exp{\left[ i k_0 \hat{n} \cdot \vec{r}_{i,j} \right]}
    \label{Eq:intensity}
\end{equation}
where $k_0 = 2 \pi / \lambda_0$,  $\vec{r}_{i,j} = \vec{r}_i - \vec{r}_j$, $\hat{n}$ is the unitary vector pointing towards the detector and $  \sigma_i^-  = \ket{g_i}\bra{e_i}$ is the lowering operator of atoms at position $\vec{r}_i$. Naturally, this quantity depends on the detection direction. On the other hand, the sum of the individual excited state populations is equal to the collective atomic state population, $\Sigma_e = \sum _ i \langle e_i \rangle = \langle \sum _ i e_i \rangle $ where $e_i = \ket{e_i}\bra{e_i}$. This is an important difference in terms of the information extracted from the system as stressed by, e.g., the protocol for entanglement witnesses presented in \cite{Ros24}, where fluctuations of the collective population and optical coherence are required. In this letter, we measure both observables in a regime where they are predicted to give the same results for a complex collective dynamics, and show that they provide the same subradiant timescales, consistent with theoretical models.

\begin{figure*}
\centering
\includegraphics[width=0.8\linewidth]{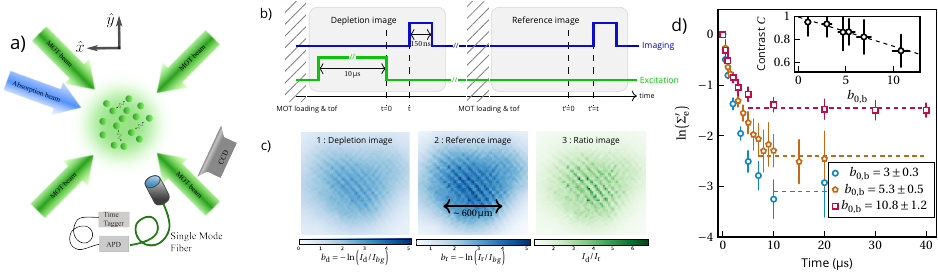} 
\caption{a) Schematic representation of the experimental  setup detailed in the main text. b) Chronogram of the experimental sequence to take depletion images. c) Example of one depleted, reference and the computed ratio image, illustrating the method. Note that the fringes in the images are residual interference effects that are not fully suppressed using a reference image, but not a fundamental feature of the atomic cloud. d) Typical experimental datasets showing the measured excited state populations normalized to their corresponding steady-state value $\Sigma'_e(t)$. The different colors and markers correspond to different optical depths as indicated in the legend of the figure. The corresponding dashed lines show the background level $\Sigma'_{e,t\rightarrow\infty}$ identified in each dataset, which is the main limitation of the method (see Supplemental materials). The inset shows the contrast of each dataset $C$ as a function of the optical depth.  The black, dashed line corresponds to a linear fit of $C$.}
\label{Fig:fig1}
\end{figure*}

The experimental setup is detailed in previous works \cite{Letellier2023,deMelo2024,Gli24} and is represented in Figure~\ref{Fig:fig1}(a). First, a magneto-optical trap (MOT) operating on the $\emph{broad}$ ${}^{1}S_0 - {}^{1}P_1$ transition of $^{174}\mathrm{Yb}$ ($\lambda_{\mathrm{b}}=\SI{399}{\nano\meter}$, $\Gamma_{\mathrm{b}}=2\pi\times\SI{29}{\mega\hertz}$) is loaded in $\sim \SI{1}{\second}$. We then transfer about $N\sim5\times10^8$ atoms into a second MOT on the \emph{narrow} ${}^{1}S_0 - {}^{3}P_1$ intercombination line ($\lambda_{\mathrm{g}} = \SI{555.8}{\nano\meter}$, $\Gamma_\mathrm{g} = 2\pi\times\SI{182}{\kilo\hertz}$). This transition allows us to reach peak optical depths up to $b_{0,\mathrm{g}} \sim 50$ even with relatively dilute atomic density up to  $\sim 10^{12}\SI{}{at\per\centi\meter\cubed}$. After this initial state preparation, the magnetic field and optical beams are switched off, and the cloud expands in free flight, which allows us to vary the peak optical depth $b_{0,\mathrm{g}}$. The optical depth is the control parameter of the collective effects in our regime of study \cite{Gue16}. We characterized our atomic clouds by absorption imaging on the broad transition, and the optical depth on the narrow transition is inferred \cite{Gli24}. 
To probe cooperativity effects, after reaching the desired optical depth, we apply a strong pulse of light on resonance with the narrow  transition, during $\sim \SI{10}{\micro\second} \gg 1/\Gamma_{\mathrm{g}}$ to ensure that a steady-state is reached. This pulse is applied using the six MOT beams. Previous works showed that the characteristic subradiant decay time does not vary with the intensity of the excitation, but the population of subradiant modes is enhanced by an optical pumping mechanism through higher excited states, beyond the limit of linear response \cite{Cip21,Glicenstein_Ferioli_Browaeys_Ferrier-Barbut_2022}. Therefore, we exploit this mechanism by choosing a total intensity of about $I \simeq I_{\mathrm{sat}} = \SI{0.14}{\milli\watt\per\centi\meter\squared}$ per beam. A similar protocol has previously been used to observe subradiant decay by measuring the photons emitted by atomic clouds \cite{Gue15,Das20,Ferioli_Glicenstein_Henriet_Ferrier-Barbut_Browaeys_2021}. In this work, we measure the characteristic subradiant decay time using both the collected scattered intensity at a given angle and the excited state population of the cloud. For the measurement of the scattered intensity, we collect the scattered photons that impinge in a solid angle $\Omega \simeq \SI{0.01}{\steradian}$ and are coupled into a single-mode fiber placed at an angle $\theta = \SI{35}{\degree}$ with respect to the $\hat{y}$ axis of the experiment (Fig \ref{Fig:fig1}). As subradiant emission is expected to be isotropic on average, this direction is arbitrary. We however avoid to be close of the 6 MOT beams to avoid direct exposure by the driving beams. The fiber is connected to a single photon counting module (SPCM-AQRH-14-FC) with a dead time of $\SI{22}{\nano\second}$ and a detection efficiency of $50\%$. This is in turn coupled to a counting card (Time Tagger Ultra, Swabian Instruments). Scattered photons are recorded for $\SI{50}{\micro \second}$ after the end of the excitation pulse allowing a continuous measurement of scattered light. We integrate many repetitions, typically for about $6-24$ hours with a duty cycle of about $\SI{1}{\hertz}$, to collect sufficient statistics depending on the exact experimental conditions.  For the measurement of the excited state population, we use the \emph{depletion} imaging technique demonstrated in \cite{Gli24}. Here we exploit the fact that alkaline-earth and alkaline-earth-like atoms such as Sr, Yb, or Dy feature two laser-cooling transitions: a broad, dipole allowed transition and a much narrower intercombination transition. The large difference in the characteristic times of each transition allows us to use the broad transition to probe the dynamics of the internal state of the atoms in the narrow transition. To monitor the time evolution of  excited state population, we take two absorption images using a beam resonant with the broad (blue, $\lambda_\mathrm{b}$) transition: the reference image $I_{\mathrm{r}}$ with no excited state population in the $^3P_1$ state and the depleted image $I_{\mathrm{d}}$ with excited state population on two clouds with similar parameters. Figure~\ref{Fig:fig1}(c) shows some typical experimental images. In a similar fashion to standard absorption imaging, we compute the ratio between the two images: $I_{\mathrm{d}}/I_{\mathrm{r}} (x,y) = \exp [ \sigma_{\mathrm{sc}}^\mathrm{b} \int{ \rho_\mathrm{e}(x,y,z) \mathrm{d}z} ] \approx 1 + \sigma_{\mathrm{sc}}^\mathrm{b} \rho_{e,\perp(x,y)} \rho_{e,\parallel}$, where $\rho_e = \rho_\mathrm{r} - \rho_\mathrm{d}$ is the spatial density of excited state atoms, and $\rho_{e,\perp}(x,y)$ the spatial density in the transverse plane, $\sigma_{\mathrm{sc}}^{\mathrm{b}}$ the scattering cross-section of the broad imaging light and $\rho_{e,\parallel} = \int \rho_e(z) \mathrm{dz}$ the integrated column density along the imaging direction. The last approximation holds well for typical experimental densities of less than $10^{12} \mathrm{at}/\mathrm{cm}^3$ and typical atomic cloud radius $r \sim \SI{400}{\micro\meter}$.
\noindent From the expression above, it is clear that we can obtain the total number of excited state atoms $\Sigma_{\mathrm{e}}$ integrating the ratio of the two images:
\begin{equation}
    \Sigma_{\mathrm{e}} =\frac{1}{\sigma_{\mathrm{sc,b}}} \int{\int{ R(x,y)}{\ \mathrm{d}x \ \mathrm{d}y }},
    \label{Eq:population}
\end{equation}
where $R(x,y)=I_{\mathrm{d}}(x,y)/I_{\mathrm{r}}(x,y)-1$. The temporal resolution of this method is given by the aborption pulse duration. In our case, a $\SI{200}{\nano\second} \ll 1/\Gamma_\mathrm{g} \simeq \SI{1}{\micro\second}$ pulse allows us to resolve the dynamics of the transition on the intercombination line while using standard optical and electronic components.

\begin{figure}[h]
    \begin{picture}(0,120)
    	\put(10,-5){\includegraphics[width=.4\textwidth]{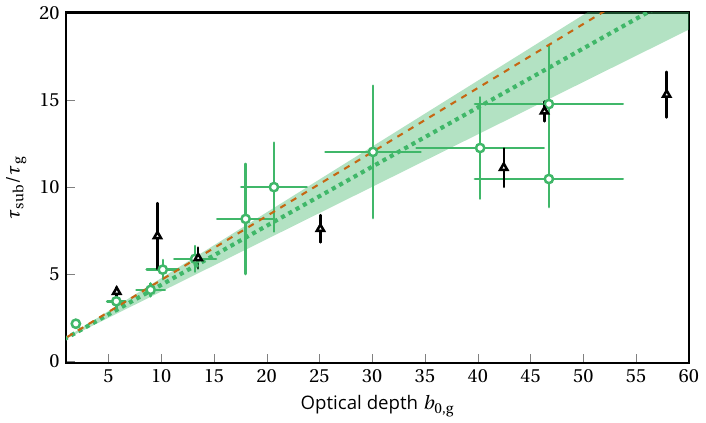}}
    \end{picture}
    \caption{Subradiance with populations and scattered light measurements. Measured late decay times as a function of the resonant optical depth $b_{0,\mathrm{g}}$. Green points correspond to measurements of the excited state population. Black triangles are the result of the scattered intensity measurements. The dashed, orange line corresponds to the numerically obtained value              $\tau_{\mathrm{sub}}/ \tau_{\mathrm{g}} \sim 0.367(9) ~ b_{0,\mathrm{g}}$. The green dotted line is the result of a linear fit of the experimental data of the population measurements giving a scaling $\sim 0.34(4) ~ b_{0,\mathrm{g}}$. The shaded area corresponds to the $\SI{68}{\percent}$ confidence error on the fit, in good agreement with theory. Vertical error bars are from the fitting procedure (see supplementary materials). The horizontal error bars on the population measurements come from sample standard deviation of the estimated optical depth during the measurements using the reference images.}
    \label{Fig:fig2}
\end{figure}

Figure \ref{Fig:fig1}(b) illustrates the chronogram for the experimental procedure to measure the excited state population. For each observation time $t$ during the decay, we prepare two clouds with the same parameters: one with the excitation beam and one without, and we image them obtaining $I_{\mathrm{d}}$ and  $I_{\mathrm{r}}$. We record at least 50 absorption images $I_{\mathrm{d,r}}$ for each experimental condition, including subtraction of the CCD dark counts and background, $I_{\mathrm{bg}}$ . The average depleted image is divided by the average reference image to produce the final experimental image, as shown in  Figure \ref{Fig:fig1}(c). The time-dependent population of excited atoms in the cloud $\Sigma_{\mathrm{e}}(t)$ is extracted using Equation \ref{Eq:population}.  Figure \ref{Fig:fig1}(d) shows typical experimental dataset showing the measured excited state populations normalized to their corresponding steady-state value 
$\Sigma'_{\mathrm{
e}}(t) = \Sigma_{\mathrm{e}}(t)/\Sigma_{\mathrm{e}}(t=0)$, for various optical depths \footnote{Throughout the text, we use the superscript $\Sigma', I'$ to indicate that the signals have been normalized to their initial state value.}. We observe that for large optical depths, the limited achievable contrast $C=(1 - \Sigma'_{e,t\rightarrow\infty})/(1 + \Sigma'_{e,t\rightarrow\infty})$ for the depletion imaging technique does not allow us to obtain reliable data for $t \Gamma > 10$ (Figure \ref{Fig:fig1}\:(d)). In absorption imaging, a reduction in the imaging contrast can arise as a consequence of the saturation of the absorption picture \cite{Hue17,Rei07}, or other sources of technical noise \cite{Ore17}. Here, density-dependent effects, e.g. light-assisted collisions or radiation pressure due to the probe light can introduce a difference in the atomic density between the two images, limiting the achievable experimental contrast and in turn limiting the observation time for a given density. It is shown in the Inset of Figure \ref{Fig:fig1}(d)  that $C$ decreases with increasing $b_0$ and that the relative value of the offset at infinite times increases with the density (see Supplemental material).

\paragraph{Subradiant scaling of decay time -} Using the two methods presented above, we extract the latest available subradiant decay time $\tau_{\mathrm{sub}}$ as a function of the peak optical depth of the samples (see Supplemental material for the description of the fitting procedure). Figure~\ref{Fig:fig2} shows the results of the extracted subradiant decay time constants for both the population and intensity measurements, showing a good qualitative agreement. We fit a linear function with a single free parameter to the results of the population measurements, finding the experimental value for the scaling of the subradiant decay as a function of the peak optical depth $\tau_{\mathrm{sub}} \sim 0.34(4) ~ b_{0,\mathrm{g}}$. This value is in good agreement with numerical calculations using a model of two-level classical coupled dipoles in the weak driving limit (see Supplemental material), from which we find a numerically extracted scaling $\tau_{\mathrm{sub}} \sim 0.367(9) ~ b_{0,\mathrm{g}}$, also in agreement with previous numerical studies \cite{Gue15}. This agreement has not been observed in previous experiments with alkaline atoms using scattered intensity measurements because of the rich internal structure of alkaline atoms \cite{Gue16}. This is not the case in the $^1S_0 \rightarrow ^3P_1$ transition in $^{174}$Yb, where all Clebsch-Gordon coefficients are equal to one, offering the quantum version of a classical dipole in a zero magnetic field.

\begin{figure}[!ht]
\includegraphics[width=0.45\textwidth]{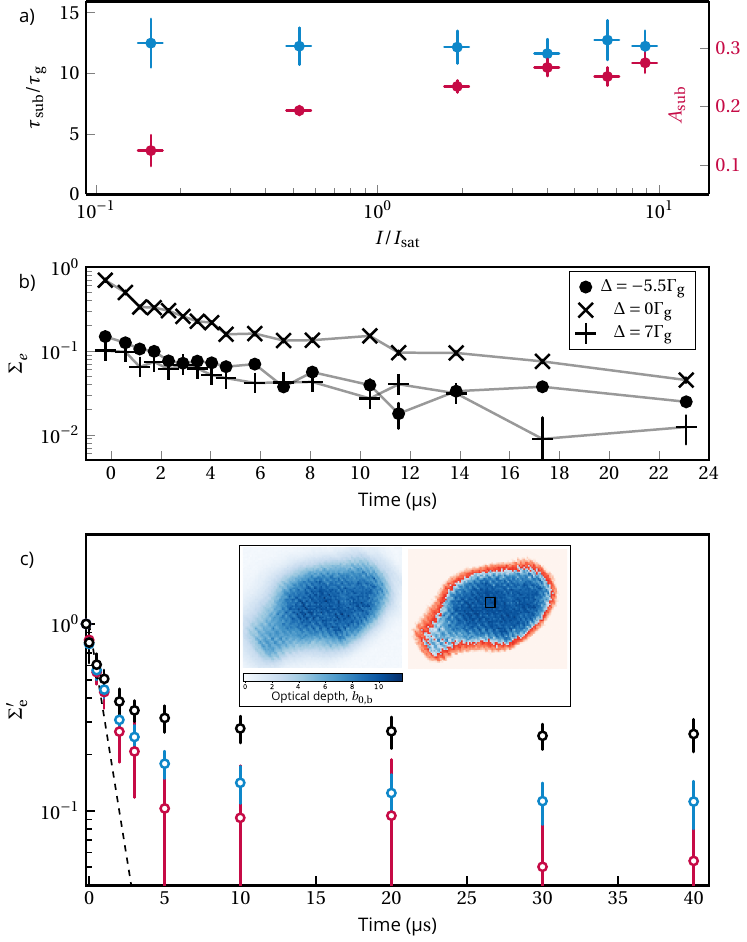}
\caption{Subradiance from population measurements. a) Blue points show the extracted latest decay time of the atomic clouds as a function of the driving intensity. Red points show the corresponding (relative) excitation that remains in the cloud after $\mathrm{t}~\Gamma_\mathrm{g} = 5$, that we define as the subradiant population. Error bars correspond to the standard error over 20 repetitions. The cloud's peak optical depth is $b_{0,\mathrm{g}} \simeq 39$. b) Normalized excited state population as a function of time for different driving frequencies. For data visualization purposes, the normalization is done to the steady-state population of the on-resonance data points. The cloud's peak optical depth is $b_{0,\mathrm{g}} \simeq 47.5$, and we measure $\tau_{\mathrm{sub}} \simeq 14 \tau_\mathrm{g}$. c) Space-resolved decay.  Points show the decay dynamics for different regions of the atomic cloud corresponding to different optical depths: black corresponds to the region with maximum optical depth in the middle of the cloud selected by the small square in the inset. Red corresponds to the region in the boundary. Blue is the average over the whole cloud. In this case, $b_{0,\mathrm{g}} \simeq 18$. The background noise depends on the selected region, in agreement with the discussion about the technical limitations of the method (See supplemental Materials).} 
\label{Fig:fig3}
\end{figure}

\paragraph{Characterization of subradiant modes using population measurements -} To emphasize the potential of the depletion imaging we present here a more detailed characterization of the subradiant modes. First, we investigate the dependence of the subradiant modes on the driving intensity. Figure~\ref{Fig:fig3}\:(a) shows the measured subradiant time constant and the subradiant amplitude $A_\mathrm{sub}$, defined as the relative population that remains after a time $\mathrm{t}~\Gamma_g  = 5$, as a function of the total driving intensity. We observe that the decay time shows no dependence on the driving intensity, while a pumping mechanism tends to populate more the subradiant modes at larger driving intensities \cite{Cip21,Glicenstein_Ferioli_Browaeys_Ferrier-Barbut_2022}. In Figure~\ref{Fig:fig3}\:(b), we show three decay curves obtained for different detunings of the probing light and with the same on-resonance optical depth.  We do not observe any dependence of the subradiant decay time constant as a function of the detuning, in agreement with previous studies \cite{Gue16,Gue17}. In addition, the depletion imaging technique, based on absorption imaging, allows us to extract a time- and spatially-resolved information of the dynamics of these systems. In this case, we resolved the time evolution of the excitations as a function of their position in the ensemble. Figure \ref{Fig:fig3}\:(c) illustrates this by considering the relative excited state population $\Sigma'_e$ for different regions of the cloud, as represented in the Inset of the figure. We observe that excitations decay faster at the boundaries, where the optical depth is lower than in the center. We attribute this to the fact that excitations are lost faster at the boundaries rather than in the center, where light is trapped by multiple scattering. While this is not surprising, we have demonstrated an experimental tool that can open the way to new investigations of the spatial structure of collective modes.

\begin{figure}[ht]
\centering
\includegraphics[width=.45\textwidth]{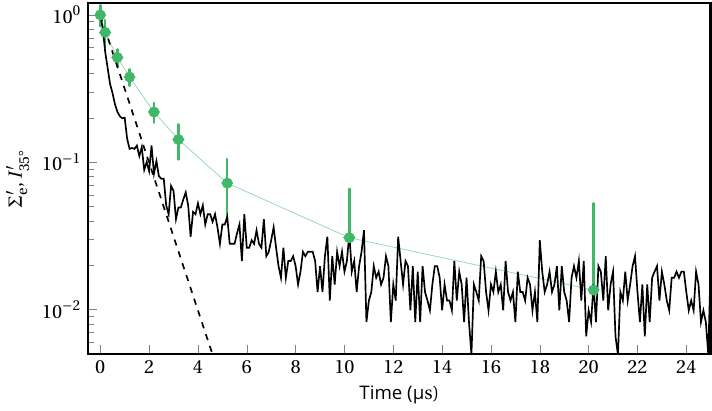}
    \caption{Black trace is the normalized intensity $I'$ collected as described in the main text. The signal corresponds to integrated photon counts in time-bins of $\delta t = \SI{0.1}{\micro\second}$. Green points are the normalized excited-state population $\Sigma`_{\mathrm{e}}$, and the line is just a guide for the eye. Dashed black line is the single-atom decay as a reference. The optical depth was $b_{0,\mathrm{g}} \simeq 24$. The two methods give the same decay rate in the subradiant regime.}
    \label{Fig:fig4}
\end{figure}

\paragraph{Differences between intensity and population measurements -} So far, we focused on how the population and intensity measurements agree regarding the measurement of the subradiant decay. However, the two observables that we are considering contain different information. In the first place, the measurement of $\Sigma_{\mathrm{e}}$ provides a  spatial resolution which is not present in the far-field emitted intensity measurements. Second, they address very different quantities, as can be appreciated from Equation~\ref{Eq:intensity} and Equation~\ref{Eq:population}. In Figure~\ref{Fig:fig4} we observe that at early times $t < 5/\Gamma_g$, the two observables behave differently due to the phase information of the individual emitters contained in the intensity measurements which is not present in the measurement of the excited state population. At later times, both observables converge because subradiant emission is incoherent. 




In this work, we present a depletion imaging method in optically dense ultra-cold $^{174}$Yb atomic samples to observe and characterize the subradiant dynamics of the system. We demonstrate the possibility to obtain spatially-resolved information during the decay and we benchmark the results with theoretical predictions and measurements of the scattered intensity. Our measurements of the subradiant decay time as a function of optical depth agree quantitatively with the theoretical predictions from numerical simulations of two-level coupled dipoles in the weak driving limit, a situation that was not previously observed before to the best of our knowledge. Our method is a powerful tool for the study of collective dynamics and can open the way to further studies of spatial correlations or spatial structure of collective modes, or obtain complementary measurements of the population dynamics in studies of collective oscillations \cite{Esp20,Hof24}.

\begin{backmatter}
\bmsection{Funding} This work was performed in the framework of the European project ANDLICA (ERC Advanced grant No. 832219), the French National Research Agency (projects PACE-
IN (ANR19-QUAN-003), LiLoA (ANR23-CE30-0035) and QUTISYM (ANR-23-PETQ-0002)). D.B.O is supported by European Union’s Horizon 2020 research and
innovation program under the Marie Skłodowska-Curie grant agreement No. 10110529.

\bmsection{Disclosures} The authors declare no conflicts of interest.

\bmsection{Data availability} Data underlying the results presented in this paper are
not publicly available but may be obtained from the authors upon reasonable
request.

\end{backmatter}

\bibliography{main_bib}
\bibliographyfullrefs{main_bib}

\end{document}